\documentstyle[subeqn,epsfig]{elsart}

\newcommand{\vect}[1]{{\rm {\bf #1}}}
\newcommand{\rmt}[1]{\tiny\rm #1}
\newcommand{\lesim}{\raisebox{-0.55ex}{$\stackrel{\displaystyle <}{\sim}$}}

\begin{document}
\begin{frontmatter}

%
\title{Analyses of two and three pion Bose-Einstein Correlations using 
Coulomb wave functions}

\author[Shinshu]{M. Biyajima\thanksref{minoru},}
\author[Shinshu]{M. Kaneyama\thanksref{masahiro} and}
\author[Toba]{T. Mizoguchi\thanksref{takuya}}
\address[Shinshu]{Department of Physics, Faculty of Science, Shinshu 
University, Matsumoto 390-8621, Japan}
\address[Toba]{Toba National College of Maritime Technology, Toba 517-8501, 
Japan}

\thanks[minoru]{E-mail: biyajima@azusa.shinshu-u.ac.jp}
\thanks[masahiro]{E-mail: kaneyama@azusa.shinshu-u.ac.jp}
\thanks[takuya]{E-mail: mizoguti@toba-cmt.ac.jp}

%
\begin{abstract}
Using effective formulas we analyze the Bose-Einstein correlations
(BEC) data corrected for Coulomb interactions provided by STAR
Collaboration and the quasi-corrected data (raw data with acceptance 
correction etc) 
on $2\pi$ and $3\pi$ BEC
by using Coulomb wave function with coherence parameter included. The
corresponding magnitudes of the interaction regions turn out to be
almost the same: $R_{\rm Coul}(2\pi) \simeq \frac 32R_{\rm Coul}(3\pi)$. 
$R_{\rm Coul}$ means the size of interaction region obtained in terms of 
Coulomb wave function. This approximate relation is also confirmed by the 
core-halo model. Moreover, the genuine 3rd order term of BEC has also been 
investigated in this framework and its magnitude has been estimated both in 
the fully corrected data and in the quasi-corrected data. 
\end{abstract}

\end{frontmatter}

%
\section{Introduction}
One of interesting subjects in heavy ion physics at high energies is
the higher order Bose-Einstein correlation(BEC) as well as the normal
(2nd order) BEC~\cite{biyajima90,csorgo99,biyajima80}. The investigation
of the genuine 3rd order term of BEC is a current important topic.
We are also interested in this subject to examine usefulness of a theoretical 
formulas described by Coulomb wave functions.
Moreover, it is well known that information on the magnitude of the interaction 
region is necessary to estimate the energy density of produced hadrons
\cite{bjorken}.  Actually several Collaborations have reported
interesting results in Refs.~\cite{na44}, \cite{wa98} and \cite{phenix}.

Very recently the STAR Collaboration~\cite{star03}
has reported new data on BEC for three negatively charged pions
(3$\pi^-$), with Coulomb corrections imposed, where $R=5$ fm is assumed as 
input~\cite{na44,star03,willson,star01}. Their data on 2$\pi$ BEC have been 
already published in
Ref.~\cite{star01}, whereas the quasi-corrected data (raw data with
acceptance correction etc) on 3$\pi^-$ BEC have been shown in 
Ref.~\cite{willson}. 
In Table~\ref{table1} we provide our classification of different
kinds of data available~\cite{bowler,biyajima96,alt,mizoguchi}\footnote{
%
%
The author of Ref.~\cite{bowler} pointed out that the Coulomb correction is 
more accurate than the Gamow correction.
}.
We shall be interested here in different kinds of data provided by the same 
experiment, to know description of them. We shall be able to examine whether 
our formulation of BEC in terms of Coulomb wave function as given in 
Ref.~\cite{mizoguchi} is useful when applied to concrete data. 

\begin{figure}[h]
  \centering
  \epsfig{file=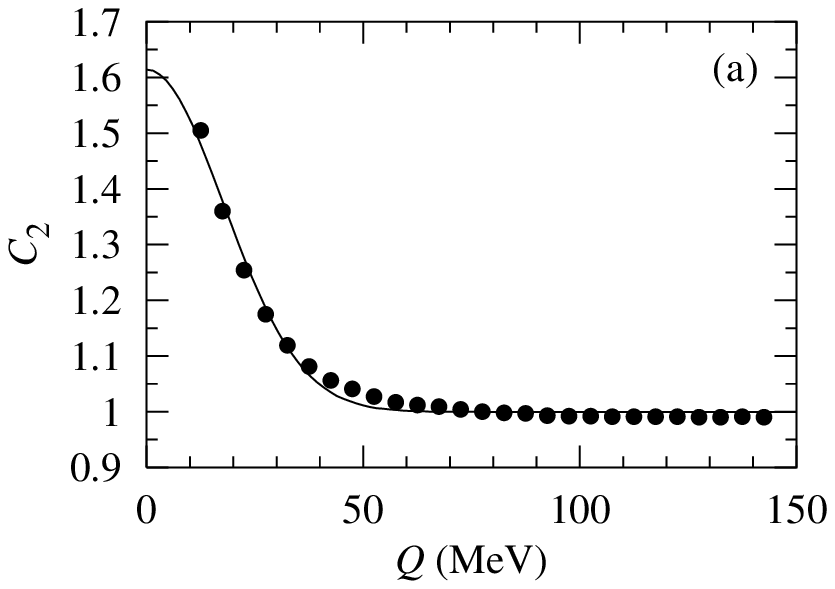,height=48mm}
  \epsfig{file=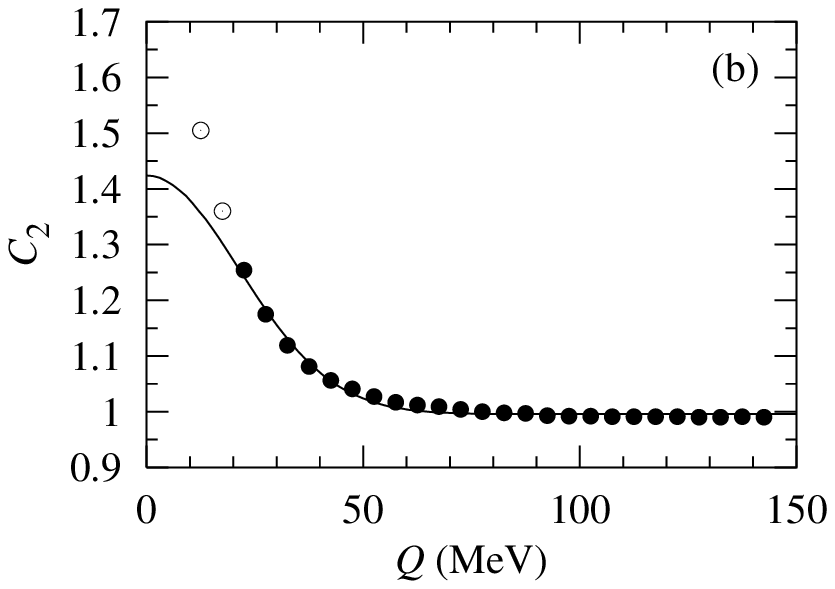,height=48mm}
  \epsfig{file=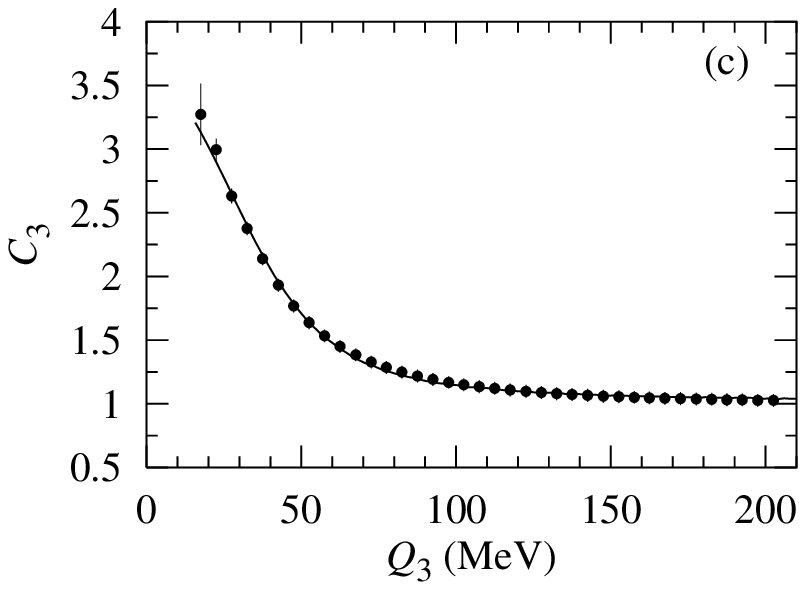,height=48mm}
  \caption{
  (a) Analysis of data~\cite{star01} by Eq.~(\ref{eq1a}). 
  (b) Data with ($\odot$) are skipped in analysis.
  (c) Analysis of data~\cite{star03} by Eq.~(\ref{eq2}).
  }
  \label{fig1}
\end{figure}

%
\begin{table}[h]
  \centering
  \caption{Category of data.}
  \label{table1}
  \begin{tabular}{ll}
  \hline
  1) Raw data  & raw data\\
  2) Quasi-corrected data (Q-CD)  
  & $({\rm raw\ data}) \times K_{\rm acceptance}$\\
  3) Gamow corrected data 
  & $({\rm raw\ data}) \times K_{\rm acceptance} \times K_{\rm Gamow}$\\
  4) Coulomb corrected data 
  & $({\rm raw\ data})  \times K_{\rm acceptance} \times K_{\rm Coulomb}$ \\
  \hline 
  \multicolumn{2}{l}{
  $K_{\rm Coulomb}$ is calculated as ${\displaystyle K_{\rm Coulomb} = 
  \frac{\int \prod d^3x_i \rho(x_i) \cdot |{\rm plane\ w.\ f.}|^2}
  {\int \prod d^3x_i \rho(x_i) \cdot |{\rm Coulomb\ w.\ f.}|^2}}$, 
  }\\
  \multicolumn{2}{l}{
  where $\rho(x_i)=(\beta^2/\pi)^{3/2} \exp[-\beta^2 x_i^2]$
  ~\cite{bowler,biyajima96,alt}. Notice that $R =$ input 
  }\\
  \multicolumn{2}{l}{
  ($\beta = 1/\sqrt 2R$) is necessary for $K_{\rm Coulomb}$, where $R = 5$ fm 
  is used in 
  }\\
  \multicolumn{2}{l}{
  Refs.~\cite{na44,star03,willson,star01}. For the systematic error bars we 
  assume 5\% for $3\pi$ BEC 
  }\\
  \multicolumn{2}{l}{
  and 0.5\% for $2\pi$ BEC.
  }\\
  \hline
  \end{tabular}
\end{table}
For various kinds of data in Table \ref{table1}, we prepare following two 
methods\footnote{
%
%
Different methods for data of $2\pi$ BEC in Refs.~\cite{adanova,sinyukov} will 
be explained in the third paragraph.
}.
\begin{description}
  \item[i) {\it A conventional method:}] Effective formulas derived from the 
  plane wave with the source function, $\rho(x_i)$ are applied to {\bf 4)} 
  Coulomb corrected data 
  categorized in Table~\ref{table1}. 
  \item[ii) {\it A method proposed in Ref.~\cite{mizoguchi}:}] A formula 
  described by the integration of the Coulomb wave function with $\rho(x_i)$ 
  is applied to {\bf 2)} quasi-corrected data(Q-CD) in 
  Table~\ref{table1}. This will be explained in the next paragraph.
\end{description}
Let us start with STAR data including Coulomb corrections 
as given by \cite{star03,star01}. We analyze these data using the following 
effective formulas, introducing parameters, $R_{\rm eff}$, $\lambda_2$ and 
$\lambda_3$,
\begin{eqnarray}
  C_2 &=& 1+\lambda_2 e^{-(R_{\rm eff}Q_{12})^2}\,,
  \label{eq1a}\\
  C_3 &=& 1+\lambda_3 \sum_{i>j} e^{-(R_{\rm eff}Q_{ij})^2}+2\lambda_3^{3/2} 
  e^{-0.5(R_{\rm eff}Q_3)^2}\,,
  \label{eq2}
\end{eqnarray}
where $Q_{ij}^2 = (\vect k_i - \vect k_j)^2$ and $Q_3^2 = (\vect k_1 - 
\vect k_2)^2 + (\vect k_2 - \vect k_3)^2 + (\vect k_3 - \vect k_1)^2$. 
 Our estimated parameters for $2\pi^{-}$ and 
$3\pi^{-}$ BEC are shown in 
Table~\ref{table3}.
To reproduce $R \approx 6.3$ fm reported by STAR Collaboration, we have to 
skip first two data points, see Fig.~\ref{fig1}(b).
%
%
%
%
\begin{table}[h]
  \centering
  \caption{Data (Coulomb correction with $R=5$ fm (input)) on $2\pi$ and 
  $3\pi$ by STAR Collaboration~\cite{star03,star01} and our analyses of data 
  by Eqs.~(\ref{eq1a}) and (\ref{eq2}).}
  \label{table3}
  \begin{tabular}{cccccc}
  \hline
  Data & Theor. & $C$ 
  & $R_{\rm eff}$ [fm] & $\lambda_2$ or $\lambda_3$ & $\chi^2/N_{\rm dof}$\\
  \hline
  2$\pi$~\cite{star01} & Eq.~(\ref{eq1a}) & 0.999$\pm$0.003 
  & 7.84$\pm$0.20 & 0.62$\pm$0.02 & 30.8/24 \\
  2$\pi$~\cite{star01} & Eq.~(\ref{eq1a}) & 0.996$\pm$0.003 
  & 6.52$\pm$0.31 & 0.43$\pm$0.03 & 8.32/22\\
  \hline 
  3$\pi$~\cite{star03,willson} & Eq.~(\ref{eq2}) & 0.999$\pm$0.012 
  & 7.83$\pm$0.26 & 0.57$\pm$0.02 & 5.32/35 \\
  \hline
  \end{tabular}
\end{table}

On the other hand, in Ref.~\cite{mizoguchi} we have proposed another 
theoretical formulation of the Coulomb wave function approach to BEC 
containing the degree of coherence parameter $\lambda$. We shall explain it 
in the next paragraph in detail. Here it will be applied to the raw data 
(with the acceptance corrections etc included) and its result will be then 
compared with the previous one in paragraph 3. Therein we shall analyze also 
the raw data on $2\pi^-$ BEC. In 4th paragraph, we are going to analyze the 
data by the core-halo model~\cite{alt,mizoguchi}. Our concluding remarks will 
be in the final paragraph. It will also contain discussion of the problem of 
phases among the three $2$nd order BEC correlation functions $C_2$, which will 
be done by introducing a parameter $\lambda$ (which can attain complex 
values). Finally we compare our results with those obtained by means of 
formula given in Ref.~\cite{heinz}.

%
\section{A theoretical formula for charged 3$\pi^{-}$ BEC.}
\begin{figure}[h]
  \centering
  \epsfig{file=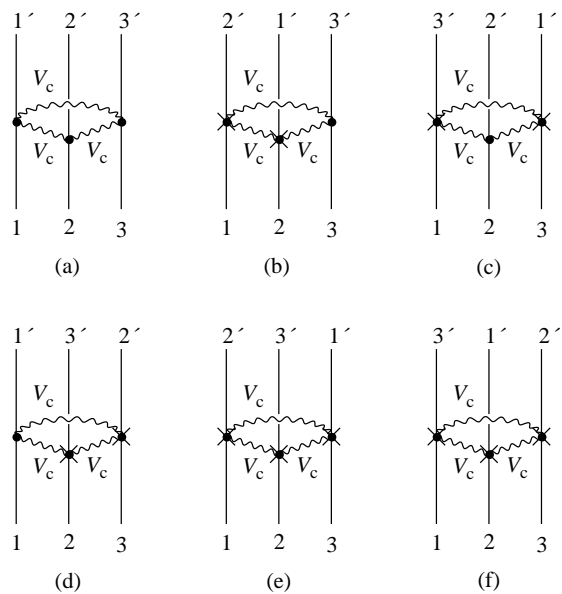,height=80mm}
  \caption{Diagram reflecting three-charged particles Bose-Einstein 
  Correlation  (BEC) and Coulomb potential ($V_c$). {\large $\times$} means 
  the exchange effect of BEC.}
  \label{fig2}
\end{figure}

We shall provide now brief explanations of our formulas (see 
Ref.~\cite{mizoguchi} for more details). The charged 3$\pi^-$ BEC should be 
described by the following products of three Coulomb wave functions, see 
Fig.~\ref{fig2}.
%
%

The theoretical formula for 3$\pi^-$BEC is expressed as
\begin{eqnarray}
  \hspace*{-0.5cm}
  \frac{ N^{(3\pi^-)}}{N^{BG}} \simeq\frac 16
  \int \prod_{i=1}^3 d^3 x_i\rho (\vect x_i) &\cdot & \left|
  \psi_{\vect k_1\vect k_2}^C(\vect x_1, \vect x_2) 
  \psi_{\vect k_2\vect k_3}^C(\vect x_2, \vect x_3) 
  \psi_{\vect k_3\vect k_1}^C(\vect x_3, \vect x_1)\right .\nonumber\\
  && + 
  \psi_{\vect k_1\vect k_2}^C(\vect x_1, \vect x_3) 
  \psi_{\vect k_2\vect k_3}^C(\vect x_3, \vect x_2) 
  \psi_{\vect k_3\vect k_1}^C(\vect x_2, \vect x_1)\nonumber\\
  && + 
  \psi_{\vect k_1\vect k_2}^C(\vect x_2, \vect x_1) 
  \psi_{\vect k_2\vect k_3}^C(\vect x_1, \vect x_3) 
  \psi_{\vect k_3\vect k_1}^C(\vect x_3, \vect x_2)\nonumber\\
  && + 
  \psi_{\vect k_1\vect k_2}^C(\vect x_2, \vect x_3) 
  \psi_{\vect k_2\vect k_3}^C(\vect x_3, \vect x_1) 
  \psi_{\vect k_3\vect k_1}^C(\vect x_1, \vect x_2)\nonumber\\
  && + 
  \psi_{\vect k_1\vect k_2}^C(\vect x_3, \vect x_1) 
  \psi_{\vect k_2\vect k_3}^C(\vect x_1, \vect x_2) 
  \psi_{\vect k_3\vect k_1}^C(\vect x_2, \vect x_3)\nonumber\\
  && \left . + 
  \psi_{\vect k_1\vect k_2}^C(\vect x_3, \vect x_2) 
  \psi_{\vect k_2\vect k_3}^C(\vect x_2, \vect x_1) 
  \psi_{\vect k_3\vect k_1}^C(\vect x_1, \vect x_3)\right|^2 .
  \label{eq3}
\end{eqnarray}
Here $\psi_{\vect k_i\vect k_j}^C(\vect x_i,\ \vect x_j)$ are the 
Coulomb wave functions of the respective 2-body collision expressed as, 
\begin{eqnarray}
  \hspace*{-0.5cm}
  \psi_{\vect k_i \vect k_j}^C(\vect x_i,\ \vect x_j) =  \Gamma(1 + i\eta_{ij})
  e^{\pi \eta_{ij}/2} e^{ i \vect k_{ij} \cdot \vect r_{ij} }
  F[- i \eta_{ij},\ 1;\ i ( k_{ij} r_{ij} - \vect k_{ij} \cdot \vect r_{ij} )],
  \label{eq4}
\end{eqnarray}
with $\vect r_{ij} = (\vect x_i - \vect x_j)$, $\vect k_{ij} = (\vect k_i - 
\vect k_j)/2$, $r_{ij} = |\vect r_{ij}|$, $k_{ij} = |\vect k_{ij}|$ and 
$\eta_{ij} = m\alpha/k_{ij}$. $F[a,\ b;\ x]$ and $\Gamma(x)$ 
are the confluent hypergeometric function and the Gamma function, 
respectively.\\

In order to introduce a parameter of the degree of coherence ($\lambda$) in 
Eq.~(\ref{eq3}), we decompose the Coulomb wave functions assigning to 
amplitudes $A(i)$ the following Coulomb wave functions (taken in our 
calculation in the plane wave approximation (PWA)):
\begin{subeqnarray}
  \label{eq5}
  A(1) &=& 
  \psi_{\vect k_1\vect k_2}^C(\vect x_1,\ \vect x_2)
  \psi_{\vect k_2\vect k_3}^C(\vect x_2,\ \vect x_3)
  \psi_{\vect k_3\vect k_1}^C(\vect x_3,\ \vect x_1)\nonumber\\
  && \stackrel{\rmt{PWA}}{\longrightarrow}
  e^{ i \vect k_{12} \cdot \vect r_{12}}
  e^{ i \vect k_{23} \cdot \vect r_{23}}
  e^{ i \vect k_{31} \cdot \vect r_{31}}
  = e^{ (3/2)i (\vect k_1 \cdot \vect x_1
          + \vect k_2 \cdot \vect x_2
          + \vect k_3 \cdot \vect x_3)}\ ,\\
A(2) &=& 
  \psi_{\vect k_1\vect k_2}^C(\vect x_1,\ \vect x_3) 
  \psi_{\vect k_2\vect k_3}^C(\vect x_3,\ \vect x_2) 
  \psi_{\vect k_3\vect k_1}^C(\vect x_2,\ \vect x_1)\nonumber\\
  && \stackrel{\rmt{PWA}}{\longrightarrow} 
  e^{ i \vect k_{12} \cdot \vect r_{13}} 
  e^{ i \vect k_{23} \cdot \vect r_{32}} 
  e^{ i \vect k_{31} \cdot \vect r_{21}}
  = e^{ (3/2)i (\vect k_1 \cdot \vect x_1
          + \vect k_2 \cdot \vect x_3
          + \vect k_3 \cdot \vect x_2)}\ ,\\
  A(3) &=& 
  \psi_{\vect k_1\vect k_2}^C(\vect x_2,\ \vect x_1)
  \psi_{\vect k_2\vect k_3}^C(\vect x_1,\ \vect x_3)
  \psi_{\vect k_3\vect k_1}^C(\vect x_3,\ \vect x_2)\nonumber\\
  && \stackrel{\rmt{PWA}}{\longrightarrow}
  e^{ i \vect k_{12} \cdot \vect r_{21}}
  e^{ i \vect k_{23} \cdot \vect r_{13}}
  e^{ i \vect k_{31} \cdot \vect r_{32}}
  = e^{ (3/2)i (\vect k_1 \cdot \vect x_2
          + \vect k_2 \cdot \vect x_1
          + \vect k_3 \cdot \vect x_3)}\ ,\\
  A(4) &=& 
  \psi_{\vect k_1\vect k_2}^C(\vect x_2,\ \vect x_3) 
  \psi_{\vect k_2\vect k_3}^C(\vect x_3,\ \vect x_1) 
  \psi_{\vect k_3\vect k_1}^C(\vect x_1,\ \vect x_2)\nonumber\\
  && \stackrel{\rmt{PWA}}{\longrightarrow}
  e^{ i \vect k_{12} \cdot \vect r_{23}}
  e^{ i \vect k_{23} \cdot \vect r_{31}}
  e^{ i \vect k_{31} \cdot \vect r_{12}}
  = e^{ (3/2)i (\vect k_1 \cdot \vect x_2
          + \vect k_2 \cdot \vect x_3
          + \vect k_3 \cdot \vect x_1)}\ ,\\
  A(5) &=& 
  \psi_{\vect k_1\vect k_2}^C(\vect x_3,\ \vect x_1)
  \psi_{\vect k_2\vect k_3}^C(\vect x_1,\ \vect x_2)
  \psi_{\vect k_3\vect k_1}^C(\vect x_2,\ \vect x_3)\nonumber\\
  && \stackrel{\rmt{PWA}}{\longrightarrow}
  e^{ i \vect k_{12} \cdot \vect r_{31}}
  e^{ i \vect k_{23} \cdot \vect r_{12}}
  e^{ i \vect k_{31} \cdot \vect r_{23}}
  = e^{ (3/2)i (\vect k_1 \cdot \vect x_3
          + \vect k_2 \cdot \vect x_1
          + \vect k_3 \cdot \vect x_2)}\ ,\\
  A(6) &=& 
  \psi_{\vect k_1\vect k_2}^C(\vect x_3,\ \vect x_2) 
  \psi_{\vect k_2\vect k_3}^C(\vect x_2,\ \vect x_1) 
  \psi_{\vect k_3\vect k_1}^C(\vect x_1,\ \vect x_3)\nonumber\\
  && \stackrel{\rmt{PWA}}{\longrightarrow} 
  e^{ i \vect k_{12} \cdot \vect r_{32}} 
  e^{ i \vect k_{23} \cdot \vect r_{21}} 
  e^{ i \vect k_{31} \cdot \vect r_{13}}
  = e^{ (3/2)i (\vect k_1 \cdot \vect x_3
          + \vect k_2 \cdot \vect x_2
          + \vect k_3 \cdot \vect x_1)}\ .
\end{subeqnarray}
Using $A(i)$,\footnote{
%
%
The normal plane wave amplitude $A_{\rm plane}(i)$ are expressed as
$$
  A_{\rm plane}(1) = e^{ i (\vect k_1 \cdot \vect x_1
                          + \vect k_2 \cdot \vect x_2
                          + \vect k_3 \cdot \vect x_3)},\ 
  A_{\rm plane}(2) = e^{ i (\vect k_1 \cdot \vect x_1
                          + \vect k_2 \cdot \vect x_3
                          + \vect k_3 \cdot \vect x_2)},\ \cdots
$$
See Ref.~\cite{csorgo99}.
}
we obtain the following three groups 
denoted as $F_i$'s,
\begin{subeqnarray}
\label{eq6}
  F_1 &=& \frac 16 \sum_{i=1}^6 A(i)A^*(i) 
\stackrel{\rmt{PWA}}{\longrightarrow} 1\ ,\\
  F_2 &=& \frac 16 [ A(1)A^*(2) + A(1)A^*(3) + A(1)A^*(6) + A(2)A^*(4) + 
A(2)A^*(5)\nonumber\\
  && \quad + A(3)A^*(4) + A(3)A^*(5) + A(4)A^*(6) + A(5)A^*(6) + {\rm 
c.\ c.}]\\
  && \quad \stackrel{\rmt{PWA}}{\longrightarrow} 
  \mbox{ BEC between two-charged particles}\\
  F_3 &=& \frac 16 [ A(1)A^*(4) + A(1)A^*(5) + A(2)A^*(3) + A(2)A^*(6) + 
A(3)A^*(6)\nonumber\\
  && \quad + A(4)A^*(5) + {\rm c.\ c.}]\\
  && \quad \stackrel{\rmt{PWA}}{\longrightarrow} 
  \mbox{ BEC among three-charged particles}
\end{subeqnarray}
We have to assign the degree of coherence ( real number of
$\sqrt{\lambda}$) to the  cross marks (x) in Fig.~\ref{fig2}.
We obtain finally that
\begin{eqnarray}
\frac{N^{(3\pi^-)}}{N^{BG}}=C 
  \int \prod_{i=1}^3 d^3 x_i\rho (\vect x_i)\cdot
  [F_1 + \lambda F_2 + \lambda^{3/2} F_3]\ ,
  \label{eq7}
\end{eqnarray}
where $\rho (\vect x)=(\beta^2/\pi)^{3/2}\exp[-\beta^2 \vect x^2]$
($\beta = 1/\sqrt 2R$) and $\lambda$ is the degree of coherence.
Moreover, 
we can also derive simple formula which uses only plane wave
approximation (PWA). It can be written as
\begin{eqnarray}
 \frac{N^{(3\pi^-)}}{N^{BG}}= {\rm Eq.\ (\ref{eq7})} 
 \stackrel{\eta_{ij} \to 0}{\longrightarrow} 
  C\left(1 + 3\lambda \sum_{i>j} e^{-\frac 94 R^2Q_{ij}^2} + 2\lambda^{3/2} 
  e^{-\frac 98 R^2Q_3^2}\right)\ ,
  \label{eq8}
\end{eqnarray}
Actually, Eq.~(\ref{eq8}) is also applied for the Gamow corrected 
data, because it contains no Coulomb effect. $R$ stands for $R_{\rm Gamow}$.

%
\section{Analyses of data on BEC.}
\begin{figure}[h]
  \centering
  \epsfig{file=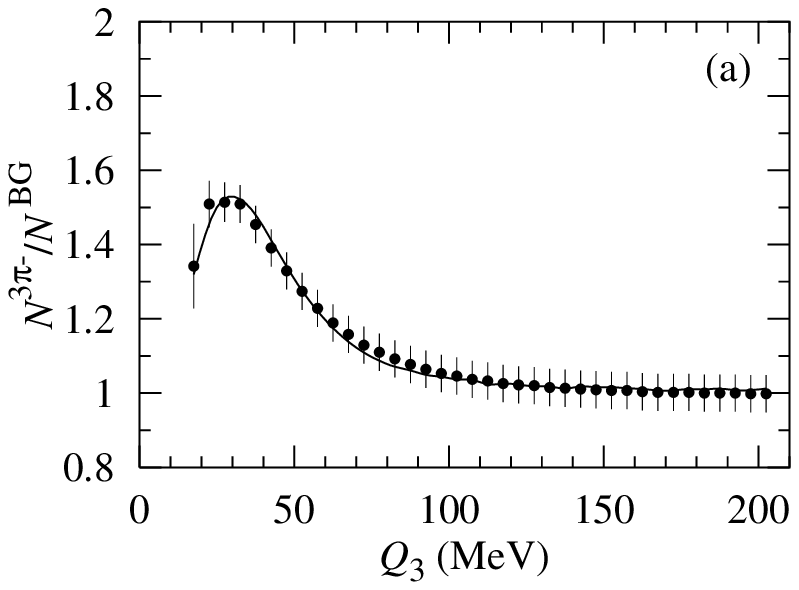,height=48mm}
  \epsfig{file=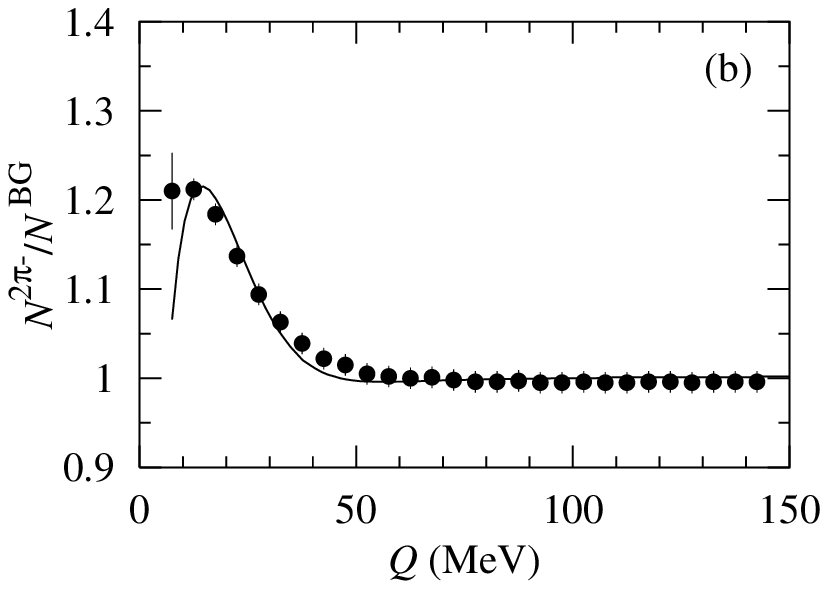,height=48mm}
  \caption{
  (a) Analysis of quasi-corrected data (Q-CD) on 3$\pi^-$BEC~\cite{star03} by 
  Eq.~(\ref{eq7}). 
  (b) Analysis of quasi-corrected data (Q-CD) on 2$\pi^-$ \cite{star01} by 
  Eq.~(\ref{eq9}).
  } 
  \label{fig3}
\end{figure}

{\bf a) Analyses of data on 3$\pi^{-}$ BEC by Eqs.~(\ref{eq7}) and 
(\ref{eq8}): } 
In Ref.~\cite{willson} quasi-corrected data at $\sqrt{s_{NN}}=130$ GeV 
have been reported. We can confirm them by using method proposed in 
Refs.~\cite{alt,mizoguchi} and results are shown in Table~\ref{table4} 
and Fig.~\ref{fig3}. 
%
%
Eq.~(\ref{eq7}) and Eq.~(\ref{eq8}) with the Gamow correction
$K_{\rm Gamow}$ are used in the present analyses. It is 
very interesting that the 
corresponding values of size parameter $R$ estimated
by Eqs.~(\ref{eq7}) and (\ref{eq8}) are almost the same. 
%
%
\begin{table}[h]
  \centering
  \caption{
  a) Analyses of two kinds of data on 3$\pi^-$ BEC. $R$ means $R_{\rm Coul}$ 
  or $R_{\rm Gamow}$. 
  b) Analyses of two kinds of data on 2$\pi^-$ BEC by means of 
  Eqs.~(\ref{eq9}) and (\ref{eq10}).
  }
  \label{table4}
  \begin{tabular}{ccccc}
  \hline
  a) 3 $\pi^-$ BEC Data &  & $R$ [fm] & $\lambda$ 
  & $\chi^2$/n.d.f.\\
  \hline
  quasi-corrected data(Q-CD) & Eq.~(\ref{eq7}) & 5.34$\pm$0.24 & 0.56$\pm$0.02 
  & 2.80/35\\
  (Q-CD)/$K_{\rm Gamow}$ & Eq.~(\ref{eq8}) & 5.03$\pm$0.20 & 0.61$\pm$0.03 
  & 6.41/35\\
  \hline
  b) 2 $\pi^-$ BEC Data &  & $R$ [fm] & $\lambda_2$ 
  & $\chi^2$/n.d.f.\\
  \hline
  quasi-corrected data(Q-CD) & Eq.~(\ref{eq9}) & 8.75$\pm$0.31 & 0.58$\pm$0.02 
  & 23.0/25\\
  (Q-CD)/$K_{\rm Gamow}$ & Eq.~(\ref{eq10}) & 7.41$\pm$0.20 & 0.61$\pm$0.02 
  & 39.4/25\\
  \hline
  \end{tabular}
\end{table}

{\bf b) Analyses of data on 2$\pi^{-}$ BEC: }
In Ref.~\cite{star01} quasi-corrected data (Q-CD) on 2$\pi^{-}$ BEC are given, 
which we have analyzed using following theoretical formula:
\begin{eqnarray}
  \frac{N^{(2+\ {\rm or }\ 2-)}}{N^{BG}} \simeq C 
  \int \prod_{i=1}^2 d^3 x_i\rho (\vect x_i)\cdot
  [G_1 + \lambda G_2]\, ,
  \label{eq9}
\end{eqnarray}
where 
$$
G_1 = \frac 12 \left (\left |\psi_{\vect k_1 \vect k_2}^C(\vect x_1,\ 
\vect x_2)\right |^2  + \left |\psi_{\vect k_1 \vect k_2}^C(\vect x_2,\ 
\vect x_1)\right |^2\right )
$$
and 
$$
G_2 = {\rm Re}\left (\psi_{\vect k_1 
\vect k_2}^{C*}(\vect x_1,\ \vect x_2) \psi_{\vect k_1 
\vect k_2}^C(\vect x_2,\ \vect x_1)\right )\ .
$$
Applying the plane wave approximation (PWA) to the above equation\footnote{
%
%
It is worth while to mention a different method for 2$\pi$ BEC utilized in 
Refs.~\cite{adanova} based on theoretical studies of Ref.~\cite{sinyukov}. The 
following formula is applied to {\bf 2)} Quasi-corrected data (Q-CD) including the momentum resolution in Pb + Au collision at 40, 80 and 158 AGeV~\cite{adanova},
$$
C_2 = C [1+\lambda' ((1+G)F^* - 1)].
$$ 
Here $G=\exp[-R_L^2 q_L^2 - R_O^2 q_O^2 - R_S^2 q_S^2 - 2R_{OL}^2q_L q_O]$. 
$R_{L}$, $R_O$, $R_S$ and cross-term $R_{OL}$ are the Gaussian source radii, 
whose source functions are assumed. $q_{\rm inv}^2 = (\vect k_1-\vect k_2)^2 - 
(E_1-E_2)^2$, and $F^* = w(k_t)\cdot (F_{\rm Coul}(q_{\rm inv})-1)+1$, where 
$F_{\rm Coul} (q_{\rm inv})$ denotes the Gaussian integration of the Coulomb 
wave function times the source function with $R = 5$ fm. See 
Ref.~\cite{adanova} for the notations, where the Monte Carlo method is also 
introduced. In this approach the effective degree of coherence 
$\lambda = \lambda' w(k_t)\cdot (F_{\rm Coul}(q_{\rm inv})-1)$ is expressed by 
the parameter $\lambda'$, the weight function $w(k_t)$ and 
$F_{Coul}(q_{\rm inv})$, where $k_t = \frac 12|\vect k_{1t} + \vect k_{2t}|$. 
Finally, physical information on $R_L$, $R_O$, $R_S$ and $R_{OL}$ are obtained 
as function of $k_t$. Various theoretical formulas with $w(k_t) = 1$ are 
investigated in Ref.~\cite{sinyukov}.
},
we obtain the following expression, 
\begin{eqnarray}
  C_2 /K_{\rm Gamow} &=& [ 1+\lambda_2 e^{-(R_{\rm eff}Q_{12})^2}]\,, 
  \label{eq10}
\end{eqnarray}
where $R_{\rm Gamow} = R_{\rm eff}$ (cf., Eq.~(\ref{eq1a})).
The results are also given in Table~\ref{table4}.
It is worth to notice that estimated values of the size parameter $R$
coincide with each other. Actually, in case of plane wave
approximation (PWA) they are reduced to the following value:
\begin{subeqnarray}
\label{eq11}
&&\frac 32R_{\rm Coul}(3\pi) \simeq  R_{\rm plane}(3\pi) \approx 
R_{\rm plane}(2\pi)\\
&&R = \frac{3}{2} \times 5\ {\rm fm} \approx 7.5\ {\rm fm}
\end{subeqnarray}
The interaction region of $R_{\rm plane}(2\pi)$ and $ R_{\rm plane}(3\pi)$ may 
be explained by the relation: 
\begin{equation}
  R({\rm Au}) = 1.2A^{\frac{1}{3}}({\rm Au}) \sim 7\ {\rm fm}\,.
  \label{eq12}\\
\end{equation}

%
\section{Analysis of data by core-halo model}
To confirm the results obtained previous paragraph, we use the core-halo model 
proposed in Ref.~\cite{csorgo99,alt,mizoguchi}. Introducing parameters
$f_c = \langle n_{\rm core}\rangle/\langle n_{\rm tot}\rangle$ (the fraction 
of multiplicity from the core part), 
$p_c = \langle n_{\rm co}\rangle/\langle n_{\rm core}\rangle$ (the fraction of 
coherently produced multiplicity from the core part), and 
$p = \langle n_{\rm chao}\rangle/\langle n_{\rm core}\rangle$ (chaoticity 
parameter defined in the laser optical approach), we have the following 
expressions with $p = 1 - p_c$,
\begin{eqnarray}
  \hspace*{-1cm}
  \frac{N^{(3-)}}{N^{BG}} &=& C \left[
  \int \prod_{i=1}^3 d^3 x_i\rho_c (\vect x_i)\cdot
  (F_1 + f_c^2p^2 F_2 + f_c^3p^3 F_3)\right .\nonumber\\
  &&\quad \left . + 
  \int \prod_{i=1}^3 d^3 x_i\rho_c (\vect x_i)\cdot \delta^3 (\vect x_2)
  \left ( 2f_c^2p(1-p) F_2 + 3f_c^3p^2(1-p) F_3 \right )\right ],
  \label{eq13}\\
  \hspace*{-1cm}
  \frac{N^{(2-)}}{N^{BG}} &=& C \left[
  \int \prod_{i=1}^2 d^3 x_i\rho_c (\vect x_i)\cdot
  (G_1 + f_c^2p^2 G_2)\right .\nonumber\\
  &&\quad \left . + 
  \int \prod_{i=1}^2 d^3 x_i\rho_c (\vect x_i)\cdot 
  \delta^3 (\vect x_2) 2f_c^2p(1-p)G_2
  \right ]\,.
  \label{eq14}
\end{eqnarray}
Here $\rho_c$ stands for the source function of the core part. The halo part 
is reflected by the delta function.

\begin{figure}[h]
  \centering
  \epsfig{file=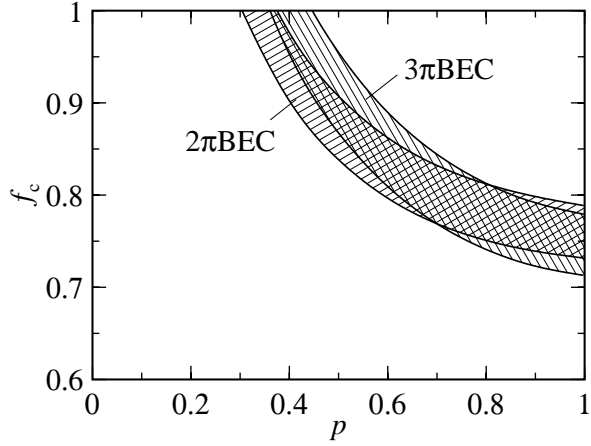,height=60mm}
  \caption{Sets of $f_c$ and $p$ estimated in analyses of BEC of 2$\pi^-$ 
  and 3$\pi^-$. Widths of $f_c$'s stand for error bar of $\pm 2\sigma$. Common 
  region are marked by mesh of a net.}
  \label{fig5}
\end{figure}

Our results are shown in Table~\ref{table6} and Fig.~\ref{fig5}. As seen in 
them, the interaction ranges of $R$ (core part) are estimated in the ranges of 
$5.3\, {\rm fm} < R_{\rm Coul}(3\pi) < 7\, {\rm fm}$ and 
$8.7\, {\rm fm} < R_{Coul}(2\pi) < 11\, {\rm fm}$, respectively. The common 
region between BEC of 3$\pi^-$  and 2$\pi^-$ is roughly described by 
$0.6\lesim p \lesim 1.0$ and $0.75 \lesim f_c \lesim 0.85$. Actually we have 
confirmed Eq.~(\ref{eq11}), $1.5R_{\rm Coul}(3\pi) \sim R_{\rm Coul}(2\pi)$.
%
%
\begin{table}[h]
  \centering
  \caption{Typical results from analysis of data by STAR Collaboration in 
  terms of Eqs.~(\ref{eq13}) and (\ref{eq14}). The effective degrees of 
  coherence are defined by $\lambda_3^* = f_c^2(p^2+2p(1-p)) + 
  f_c^3(p^3+3p^2(1-p))$ and $\lambda_2^* = f_c^2(p^2+2p(1-p))$, respectively. 
  Note that figures at $p=1$ reproduce previous results in Table~\ref{table4}.}
  \label{table6}
  \begin{tabular}{cccc}
  \hline
  $p$ & 1.0 & 0.8 & 0.6\\
  \hline
  \multicolumn{4}{c}{$3\pi$ BEC}\\
  $f_c$                & 0.75$\pm$0.02 & 0.78$\pm$0.02  & 0.85$\pm$0.02\\
  $\lambda_3^*$        & 0.97          & 1.00           & 1.00\\
  $R$ (fm)             & 5.34$\pm$0.24 & 5.99$\pm$0.28  & 6.48$\pm$0.32\\
  $\chi^2/N_{\rm dof}$ & 2.80/34       & 1.39/34        & 0.98/34\\
  \multicolumn{4}{c}{$2\pi$ BEC}\\
  $f_c$                & 0.76$\pm$0.01 & 0.78$\pm$0.02  & 0.83$\pm$0.02\\
  $\lambda_2^*$        & 0.58          & 0.59           & 0.58\\
  $R$ (fm)             & 8.75$\pm$0.31 & 10.16$\pm$0.38 & 11.12$\pm$0.42\\
  $\chi^2/N_{\rm dof}$ & 23.0/24       & 16.3/24        & 16.1/24\\
  \hline
  \end{tabular}
\end{table}
%
%

%
\section{Concluding remarks}
Because quasi-corrected data at $\sqrt{s_{NN}}=130$ GeV have been reported by 
STAR Collaboration, we have applied theoretical formula, Eq.~(\ref{eq7}) to 
analyze them. Our method and the conventional method for the charged BEC are 
explained in paragraph 1. From our results, it can be said that both methods 
are (almost) equivalent to each other (when accounting for the Fourier factor 
(3/2) in Eq.~(\ref{eq5}) in case of charged $3\pi^-$ BEC).  From our present 
study we have known that magnitude of the interaction region of Au + Au 
collisions at $\sqrt{s_{NN}}=130$ GeV is described by the relation 
$R_{\rm Coul}(2\pi) \simeq \frac 32R_{\rm Coul}(3\pi)$ whose magnitude is 
explained by $R({\rm Au}) = 1.2\times A^{1/3}$ fm. This fact is also confirmed 
at $p=1$ in Table~\ref{table6}, provided that the core-halo model proposed in 
Refs.~\cite{csorgo99,alt,mizoguchi} is used. This kind of study should be 
applied also to other data at SPS and RHIC energies~\cite{biyajima04}.

Finally we consider problem of the phase factors appearing between three 2nd 
BEC $C_2 ({\bf k}_1,{\bf k}_2)$, $C_2 ({\bf k}_2,{\bf k}_3)$, and 
$C_2 ({\bf k}_3,{\bf k}_1)$. According to Ref.~\cite{heinz}, we analyze 
the respective data by the following formula for the 3rd order BEC, 
\begin{eqnarray}
  C_3 = C[1 + \lambda_3 \sum_{i>j} e^{-(R_{\rm eff}Q_{ij})^2}
            + 2\lambda_3^{3/2} e^{-0.5(R_{\rm eff}Q_{3})^2}\times W]\,,
  \label{eq15}
\end{eqnarray}
where $ W = \cos (\phi_{12} + \phi_{23} + \phi_{31}) $, which, in the simplest 
form, is parameterized as $ W = \cos(g \times Q_3)$. 

\begin{figure}[h]
  \centering
  \epsfig{file=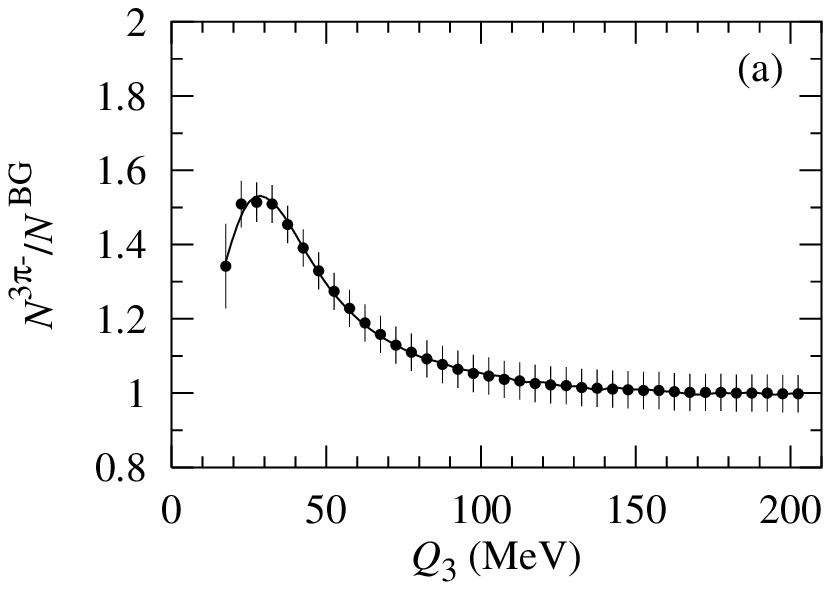,height=60mm}
  \epsfig{file=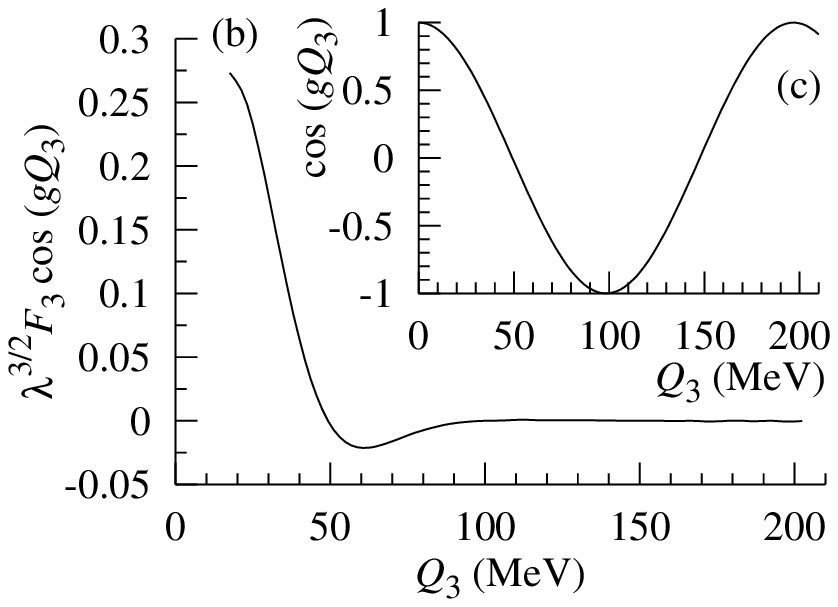,height=60mm}
  \caption{
  (a) Analysis of Q-CD by Eq.~(\ref{eq16}). 
  (b) The magnitude of the genuine term. (c) Behavior of $\cos(g\times Q_3)$.
  }
  \label{fig6}
\end{figure}

Similar formula can be obtained by replacing in Fig.~\ref{fig2} and 
Eq.~(\ref{eq8}) degree of coherence parameter $\sqrt{\lambda}$ by three 
complex numbers ($\sqrt{\lambda}e^{i\varphi}$):
\begin{eqnarray}
\frac{N^{(3\pi^-)}}{N^{BG}}=C 
  \int \prod_{i=1}^3 d^3 x_i\rho (\vect x_i)\cdot
  [F_1 + \lambda F_2 + \lambda^{3/2} F_3 \times W]\,. 
  \label{eq16}
\end{eqnarray}
Our results are shown in Table~\ref{table7} and Fig.~\ref{fig6}. The genuine 
$3$rd term of BEC in the quasi-corrected data (Q-CD) is explicitly shown in 
Fig.~\ref{fig6}. It should be stressed that the phase factor is playing 
important role in formula $N^{(3\pi^-)}/N^{BG}$~\cite{hove}, leading to 
smaller minimum of $\chi^2$, see Table~\ref{table7}. The smaller $\chi^2$'s in 
Table~\ref{table7} than those of Table~\ref{table6} suggest us the usefulness 
of Eq.~(\ref{eq16}). To elucidate this fact, study on $g$ is also necessary in 
a future~\cite{biyajima04}\footnote{
%
%
There is a possible interpretation as $g = 30\ {\rm GeV}^{-1} \simeq 6\ {\rm fm}$, provided that $0.2\ {\rm GeV}\cdot {\rm fm} = 1$ is used. 
}.
%
%
\begin{table}[h]
  \centering
  \caption{Analyses of 3$\pi^-$ BEC by means of Eq.~(\ref{eq15}) and 
  Eq.~(\ref{eq16}) with $W = \cos (g\times Q_3)$.}
  \label{table7}
  \begin{tabular}{cccccc}
  \hline
  Data &  & $R$ [fm] & $\lambda_3$ or $\lambda$ & $g$ [GeV $^{-1}$] 
  & $\chi^2$/n.d.f.\\
  \hline
  Coulomb CD & Eq.~(\ref{eq15}) & 7.08$\pm$0.42 & 0.62$\pm$0.03 & 32.9$\pm$6.2 
  & 1.00/34\\
  \hline
  (Q-CD) & Eq.~(\ref{eq16}) & 4.91$\pm$0.36 & 0.61$\pm$0.04 & 31.9$\pm$8.8 
  & 0.57/34\\
  \hline
  \end{tabular}
\end{table}
%
%

%
\section*{Acknowledgements}
One of authors (M.~B.) is indebted for critical comments concerning our 
paper~\cite{biyajima90} by late L.~Van Hove. We are also owing to G.~Wilk for 
his careful reading the manuscript. Various communications with K.~Morita, T.~Sugitate, Y.~Miyake and L.~Rosslet are also acknowledged.

\end{document}